\documentclass[useAMS,usenatbib]{mn2e}

\usepackage{times,epsfig,natbib,amssymb,amsmath,graphics,longtable,threeparttable}

\newcommand{\um}{\mbox{${\mu}$m}}             

\def \lsun{\ifmmode{{\rm\ L}_\odot}\else{${\rm\ L}_\odot $}\fi}

\def \msun{\ifmmode{{\rm\ M}_\odot}\else{${\rm\ M}_\odot$}\fi}

\def \rsun{\ifmmode{{\rm\ R}_\odot}\else{${\rm\ R}_\odot$}\fi}

\newcommand{\kms}{kms$^{-1}$}                         

\def \mdot{\ifmmode{{\rm\dot{M}}}\else{${\rm\dot{M}}$}\fi}

\newcommand{\ha}{H$\alpha${}}

\title[SNe in Arp 299]{On the multiple supernova population of Arp 299: constraints on
progenitor properties and host galaxy star formation characteristics\thanks{Based 
on observations made with the Isaac Newton Telescope and the William Herschel Telescope
operated on the island
of La Palma by the Isaac Newton Group in the Spanish Observatorio del Roque de los
Muchachos of the Instituto de Astrofisica de Canarias}}
\author[Anderson et al.]{J. P. Anderson$^{1}$\thanks{E-mail:
anderson@das.uchile.cl}, S. M. Habergham$^{2}$ \&\ P. A. James$^{2}$\\
$^{1}$Departamento de Astronom\'ia, Universidad de Chile, Casilla 36-D, 
Santiago, Chile\\
$^{2}$Astrophysics Research Institute,
Liverpool John Moores University,
Twelve Quays House,
Egerton Wharf,
Birkenhead,
CH41 1LD,
UK}

\begin{document}

\date{}

\pagerange{\pageref{firstpage}--\pageref{lastpage}} \pubyear{2010}

\maketitle

\label{firstpage}

\begin{abstract}
Arp 299 is an interacting system of two components; NGC 3690 and IC 694. Throughout the last 
20 years 7 
supernovae have been catalogued as being discovered within the system. One of these is 
unclassified, leaving 6 core-collapse supernovae; 2 type II (one with IIL sub-type 
classification), 2 type Ib events, a type IIb supernova and one object of indistinct type; 
Ib/IIb.\\
We analyse the relative numbers of these supernova types, together with their relative
positions with respect to host galaxy properties, to investigate implications
for both progenitor characteristics and host galaxy star formation properties.\\
Our main findings are: 1) the ratio of `stripped envelope' supernovae (types Ib and IIb) to 
other `normal' type II is 
higher than that found in 
the local Universe. There is $\sim$10\%\ probability that the observed
supernova type ratio is drawn from an underlying distribution such as that
found in galaxies in the local Universe. 2) All `stripped envelope' supernovae 
are more centrally concentrated within the system 
than the other type II ($\sim$7\%\ chance probability). 
3) All supernova environments have similar derived metallicities and there are no 
significant metallicity gradients found across the system. 4) 
The `stripped envelope' supernovae all fall on regions of \ha\ emission while the other type II 
are found to occur away from bright HII regions (again,
  $\sim$7\%\ chance probability).\\
From this investigation we draw two different -- but non-mutually exclusive -- interpretations
on the system and its supernovae: 1) The distribution of supernovae, and the relatively 
high fraction of types Ib and IIb events over other type II can be explained by the 
young age of the most recent star formation in the system, 
where insufficient time has expired for the 
observed to match the `true' relative supernova rates. If this explanation is
valid then the present study provides additional (\textit{independent}) evidence that both 
types Ib and IIb supernovae arise from progenitors of shorter stellar lifetime
and hence higher mass than other type II supernovae. 
2) Given the \textit{assumption} that types Ib and IIb trace 
higher mass
progenitor stars \citep{and08}, the relatively high frequency of types Ib and IIb to other
type II, and also the centralisation of the former over the latter with respect
to host galaxy light implies that in the centrally peaked and enhanced star formation within
this system, the initial mass function is biased towards the production of high mass stars.\\

\end{abstract}


\section{Introduction}
\label{intro}
Constraints can be made on supernova (SN) progenitors through studying the
host galaxies and the environments within those host galaxies
where different SN types are found. Usually this is achieved through analysis
of large
samples of SNe occurring within different galaxies. In a series of recent
papers (e.g. \citealt{and08,hab10,and10}; AJ08, H10 and A10 henceforth) 
we have drawn various
conclusions on progenitor characteristics and star formation (SF) properties
of host galaxies through studying the nature of the galaxy
light found in close proximity to historical SNe.\\ 
A small
number of galaxies have hosted multiple SNe. One such example is Arp 299 
which over the last two decades has been host to 7 detected SNe\footnote{The
`SN 1990al' designation was originally given to a possible radio SN
in Arp 299 (\citealt{hua90}). However, this has subsequently been
found to be consistent with a background PG quasar (\citealt{ulv09}), 
hence we do not include this object in the current study. Note that 
this object has now also been removed from the IAU SN 
catalogue}. 
This provides an opportunity to
study the distribution of SN types with respect to host 
galaxy characteristics with freedom from the selection effects involved in
assembling a sample of galaxies with differing properties (e.g. distance,
surface brightness and star formation histories).\\
Hence, in the current paper we use Arp 299 and its SNe as a case study to explore
the properties of SN progenitors and host galaxy SF by investigating the
relative SN numbers and the positions at which they
are found within Arp 299, both with respect to each other and with
respect to the system components.\\
The paper is organized as follows. In the next section we summarise the
overall properties of Arp 299, the extreme SF within the system, the multiple SN population, and
our current understanding of SN progenitors. Next, in section 3
we summarise the observations obtained and used for the current analysis.
In section 4 we summarise the analysis techniques used throughout the paper.
In section 5 we present our results, followed by a discussion of
the implications of these in section 6. Finally we list our
main conclusions from this work.

\section{Arp 299 and its SNe}
\label{arp299}
Arp 299 is a merging event of two galaxies; IC 694 and NGC 3690
\footnote{We note here that there appears to be some confusion between the literature and
the NED database on the naming and coordinates of Arp
299 and its components. Here we are consistent with the literature cited when
referring to different regions. This means that the values for the system we quote here
are actually those that are quoted in NED when one searches for NGC 3690}. Due to
this merging process the system is going through a period of intense
merger-driven burst of SF (e.g. \citealt{wee72,rie72,geh83}). The system lies at a distance of 43.9 Mpc with a
heliocentric recession velocity of 3121\kms (values taken from the NASA
Extragalactic Database\footnote{http://nedwww.ipac.caltech.edu/}). 
While both galaxies within the
system are heavily disturbed (see Figure 1), the Hyper-Leda
database\footnote{http://leda.univ-lyon1.fr/} lists both as
SBm morphologies, i.e. irregular barred spirals. Arp 299 has been studied
as a typical merging system extensively with observations published from radio
through to X-rays. Here we use our own observations of the system, plus where
applicable observations and measurements taken from the archives/literature in order
to further understand the galaxy system with respect to the SNe that
it has produced.

\begin{figure}
\includegraphics[width=8.5cm]{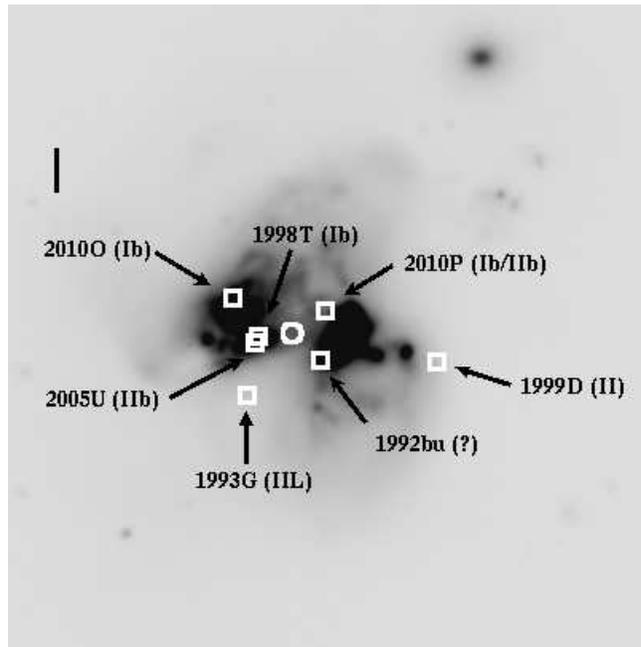}
\caption{$R$-band image of Arp 299. The eastern 
structure (to the left of the figure) is known as IC 694, while the western
structure is NGC 3690. The positions of all catalogued SNe are marked on the 
figure using squares, with SN type classification shown in brackets. The scale bar is 2 kpc in length. 
The image is orientated with east to the left and north up. The circle indicates the system
centre as listed in NED.}
\end{figure}

\subsection{The extreme star formation in Arp 299}
\label{exSF}
The numerous starburst regions within Arp 299 were originally
studied by \cite{geh83}, who compared its properties in the optical, 
infrared and radio, identifying a number of regions in the system
relating to different starburst components. These naming conventions
have continued (and been complemented with further regions for study)
in more recent work using spectroscopy and higher resolution (HST)
imaging (e.g. \citealt{alo00,gar06,alo09}). In later sections of this paper
we compare our measurements/analysis to these earlier works, in order
to check the validity of our results, and to help understand the system and its SNe.\\
It is obvious from Fig. 1 that the system is going through a strong interaction
and merging event. Arp
299 shows a single tidal tail, indicative of a prograde-retrograde
interaction initiated $\sim$700 Myr ago \citep{nef04}. While there is likely 
to be long-term
interaction induced SF within this system, \cite{alo00} have 
identified very young starburst regions within Arp 299 that
have extremely high SF rates and ages as young as 4 Myr. This latter
issue is key to some of the subsequent discussion below
when comparing SNe found in Arp 299
to the SF properties of the system. This will be discussed in detail later,
however here we make the point that even if these extreme starbursts dominate
the current SF in the system, it is highly probable that
there are SF episodes of considerably older ages which were 
induced earlier in the interaction of the two galaxies (note that even 
prior to the interaction, both of the component galaxies were spiral types, i.e.
galaxies containing significant amounts of SF).\\

\subsection{Supernovae discovered within Arp 299}
\label{sup_arp299}
During the past 20 years 7 SNe have been discovered within Arp 299. 
The positions of these SNe are shown in Fig. 1 along 
with the type classifications. 
Searching the Asiago\footnote{
http://graspa.oapd.inaf.it/} and 
IAU SN
catalogues\footnote{http://www.cbat.eps.harvard.edu/lists/Supernovae.html}, 
the first reported discovery we find is that of SN 1992bu.
SN 1992bu was discovered by \cite{van94} as part of a SN search in 
starburst galaxies using $K$-band images. 
This object has no type classification, but for completeness we include the SN
in our analysis. 
The third reported event is SN 1993G \citep{tre93}, discovered as part of the 
Leuschner Observatory Supernova Search (LOSS, the precursor to the current Lick Observatory Supernova Search).
This SN was classified as a type II SN (SNII henceforth) by \cite{fil93_2}, with a later sub-type classification
of SNIIL \citep{tsv94}, where the L signifies the `linear' classification and 
indicates that in addition to containing hydrogen in its spectrum the 
SN displayed a linear decline in its light curve (e.g. \citealt{bar79}). 
SN 1998T was discovered by the Beijing Astronomical
Observatory Supernova Survey (BAOSS) and confirmed by the Lick Supernova survey 
\citep{li98}. It was classified as a SNIb \citep{li98}, i.e. hydrogen was
lacking in the spectrum but helium was present (see \citealt{fil97}, for a review of
spectral SN classifications). In the following year SN 1999D was
discovered, again by BAOSS \citep{qiu99}, and classified as a
SNII \citep{jha99} with no further sub-type classification given. SN 2005U
was discovered by the Nuclear Supernova Search \citep{mat05}, and originally classified as
a SNII by \cite{mod05}, with a later refinement to SNIIb classification \citep{leo05}. 
SNIIb are thought to be transitional objects being similar to SNII at early times (prominent
hydrogen lines), while becoming similar to SNIb at later times \citep{fil93}.
Most recently 2 SNe were discovered at the 
start of 2010. SN 2010O was found by the Puckett Observatory Supernova Search 
\citep{new10}, and classified as a SNIb by \cite{mat10}. Somewhat surprisingly, the 
next SN listed on the IAU database, SN 2010P, was also discovered in the interacting system. 
SN 2010P was found using near-IR images by \cite{mat10_2}, and was later classified \citep{ryd10}
as SNIb/IIb.\\

We note here that in the current study we only discuss those SNe that have been
discovered using optical and near-IR galaxy images and where definitive classification as
true SNe has been possible. While we include one unclassified SN, our conclusions arise from 
those with classification into various CC types which enables us to infer
characteristics of the progenitor population and the underlying SF properties.
However, the term `supernova factory' has been coined for this galaxy system through imaging
of the system at radio wavelengths. \cite{nef04} and \cite{per09} (also see \citealt{ulv09})
both concluded from radio observations
that there are a number of sources which are likely to be recently exploded CC SNe, and
that the nucleus of the component IC 694 is thus a `supernova factory', consistent 
with the presence of very young extreme SF regions as proposed by \cite{alo00}. \cite{nef04}
conclude that the detected radio emission of the system implies supernova
rates of 0.1-1 yr$^{-1}$
(recently \citealt{rom11} also published a lower limit for the SN rate of
  0.28yr$^{-1}$ from VLA monitoring of the radio emission of the system).
Due to the high level of extinction measured in the system (possibly an $A$$_{\textit{v}}$ 
of $\sim$ 30 magnitudes in extreme cases; \citealt{alo00}), it is very probable that
many of these SNe are missed in optical/near-IR searches. If this extinction preferentially
obscures a certain type of event from discovery over others then this may affect the 
current analysis. We come back to this issue later and discuss it as a possible caveat to our
conclusions.

\subsection{Core-collapse supernova progenitors}
\label{progCC}
Core-collapse (CC) SNe are generally classified into two broad groups. Those
that show signs of hydrogen in their spectra; the type II SNe, and those that do not; the
type Ibc (throughout this paper type `Ibc' SNe refers to all those SNe
classified as SNIb, SNIc or SNIb/c in the literature). This lack of hydrogen in the latter indicates
that their progenitor stars must have lost their envelopes in their pre-SN stellar evolution through
some process. Whether this emerges from higher mass/metallicity progenitors producing
stonger stellar winds, or from binary interactions is still strongly debated. We will
not review all of the literature here (see \citealt{sma09b} for a recent review of
CC SN progenitors), but merely note the work that is most relevant to the current discussion
on differences in progenitor characteristics.
SNIIP, the most common CC SN type (observed SN rates will be discussed later)
are the only class where a number of progenitor
stars have been identified on pre-explosion images.
\cite{sma09} presented an analysis of direct detections 
and concluded that SNIIP arise from red-supergiant progenitors between 8.5 and 16.5\msun\
(we note the interesting fact that none of the 7 SNe in the current galaxy
is classified as IIP). Direct detections of other
CC SN types have been few. One of the type II in the current study is a IIL, where
the lack of a plateau (found in the IIP), in their light curves is thought to
indicate a smaller envelope remaining at the epoch of explosion. One direct detection has
been made of a progenitor of a SNIIL, with derived progenitor masses ranging from
11-24\msun\ (given the errors from the two studies: \citealt{eli10,fra10}), hence
possibly indicating that these SNe arise from slightly higher mass objects than the IIP
(however, given a sample of 1, generalising to the whole population is perhaps premature).
The IIb class has had two detections of progenitor stars on pre-SN images. \cite{cro08} concluded
that SN 2008ax either arose from a single massive star of $\sim$28\msun\ or a lower
mass binary system, while for the extensively studied SN 1993J, \cite{mau04} estimated that
the SN arose from an interacting binary with component stars of 14 and 15\msun. Therefore from
this small sample the detections suggest that the SNIIb arise from more massive progenitors
(whether single or binary star 
progenitors) than the SNIIP (we also note the likely detection of a massive binary 
companion star to the SNIIb 2001ig claimed by \citealt{ryd06}).   
A direct detection of a SNIbc has thus far eluded us \citep{sma09b}, and while upper limits from searches can 
be put on progenitors, the uncertain nature of current modeling of WR stars (the possible single 
star progenitors of SNIbc; e.g. \citealt{gas86}), means that constraints are less reliable than those
for SNIIP. \\
Other, more indirect constraints can be obtained
through studying the nature of the host galaxy emission in close proximity to SN positions. In AJ08
it was shown that SNIbc show a higher degree of asssociation than the SNII population
to host galaxy \ha\ emission, implying
shorter stellar lifetimes and therefore higher progenitor masses (see also \citealt{kel08} for
similar results). In the same study the SNIIb
also accurately traced the underlying emission (while the SNIIP did not), therefore 
again implying higher mass progenitors than normal SNII. The SNIIL showed a similar
degree of association to the emission as SNIIP implying similar mass progenitors. 
It therefore seems that
progenitor mass is playing a significant role in determining SN type, 
as suggested by single star models (e.g. \citealt{heg03,eld04,geo09}).\\
Progenitor metallicity could also be playing a role in determining SN type.
The general consensus seems to be that
the SNIbc 
should arise from higher metallicity progenitors, as this would provide assistance
in removing the outer hydrogen envelope, bringing down the progenitor mass at which SNIbc are produced. (We note 
with reference to the current study that stellar models also generally predict that SNIIb arise from
higher metallicity progenitors than SNIIP; e.g. \citealt{heg03}).
Observationally this was originally suggested by the centralisation of SNIbc within 
their host galaxies compared to SNII (e.g. \citealt{bar92,ber97,tsv04,hak09,and09}; AJ09 henceforth). However, the 
cause of this centralisation has been complicated by work separating host galaxies 
by signs of interaction/disturbance. In H10 these radial differences were found to be much more
prominent in interacting/disturbed galaxies and it was concluded that this was an effect
of top-heavy IMFs in the central regions of these galaxies. 
SNIbc are also found to occur in more luminous
(and therefore probably more metal rich) host galaxies \citep{boi09}, while measurements of the
global host galaxy metallicities of SNe \citep{pri08b} found that SNIbc again appeared to arise from higher
metallicity hosts. However, in A10, where the gas-phase metallicities of the environments
of CC SNe were derived, while a difference was found between the SNIbc and SNII, this difference
was found to be small, suggesting that metallicity does not play a significant role in determining
SN type (we also note similar studies that concentrated on metallicity
differences between SNIb and SNIc environments; \citealt{mod11,lel11}). Finally, we note that
both modeling and observations show that even the highest metallicity
environments
produce substantial numbers of SNII.\\
The above discussion shows that while much work has been done in order to further
constrain and understand the properties of SN progenitors, disentangling all the different
evidence is difficult. The present study attempts to further our understanding
by using Arp 299 and its relatively large SN population as a case study, while also using
the knowledge implied from the above research to try to further understand the
SF properties of the host system. 

\section{Observations}
\label{obs}
We obtained our own $R$-band and \ha\ imaging of Arp 299 using the Wide Field 
Camera (WFC) on the Isaac Newton Telescope (INT). These data and their reduction
have been discussed elsewhere (AJ08). Briefly, the $R$-band image was scaled to 
the \ha\ image and used to continuum subtract the latter, leaving only 
the emission from HII regions within the system. 
We also obtained long slit
optical spectra of the system, data that was initially presented in 
A10 (see here for more details of instrument setup, reduction procedures etc). 
These were obtained with the Intermediate dispersion Spectrograph and Imaging
System (ISIS) on the William Herschel Telescope (WHT).
The slit positions across the system are shown in Fig. 2. 
From the literature we use \textit{Spitzer} 24\um\ 
imaging\footnote{downloaded from http://ssc.spitzer.caltech.edu/spitzerdataarchives/}, 
plus near- and far-UV GALEX imaging\footnote{obtained from
http://galex.stsci.edu/GalexView/}. 
These images are
shown in Fig. 3 together with our own $R$-band and \ha\ imaging. Here we also show
a $K$-band image taken from the literature \citep{cal97}. Unlike the other literature images above
we do not use the near-IR image in our analysis, but include it in
Fig. 3 to show the morphology of this system at different wavelengths.

\begin{figure}
\includegraphics[width=8cm]{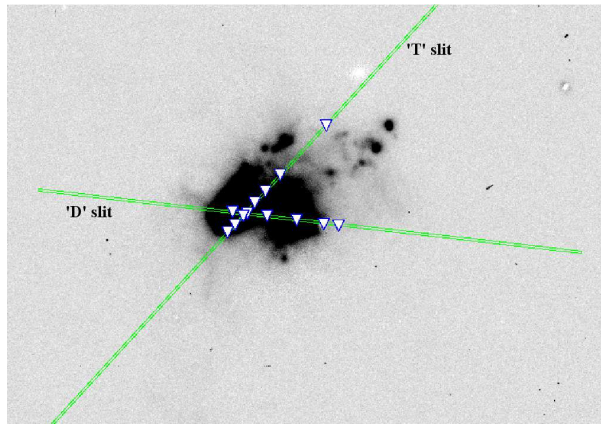}
\caption{\ha\ continuum subtracted image of the system Arp 299. 
The two lines show the positions and sizes (1 arcsecond width, 3.7
arcminute length) of the slits that were used to obtain optical spectroscopy of Arp 299.
The white triangles indicate
the extraction positions for the spectra used to measure any possible metallicity gradients
across the system. The image is orientated with north up and east to the left. In the
subsequent analysis/discussion the slit positioned at an angle of 131.5 degrees east of north
is referred as the `T' slit, while the other is referred to as the `D' slit,
as indicated in the figure.}
\end{figure}

\section{Analysis techniques}
\label{anal}
The main analysis techniques used throughout this work are those
detailed in AJ08, AJ09, and A10. Here we briefly summarise those statistical 
methods but we refer the reader to those references for a full discussion of the
methods and associated errors. We note that for three SNe
these statistics (at least with respect to the \ha\ and $R$-band emission
of their host galaxies) have already been presented in previous papers; SN 1993G, SN 1998T and
SN 1999D, but for the three most recent SNe (SN 2005U, SN 2010O and SN 2010P, plus the 
unclassified SN 1992bu) we 
present new analysis (note that SN 2005U had been discovered at the time of publication
of those previous works, however in the catalogues its host galaxy is listed as Arp 299
and not NGC 3690 as were all others, and therefore the SN was missed).\\
In \cite{jam06} we introduced a statistic which normalises the count found within
one pixel of a galaxy of some light emission (in that case \ha), to the
overall counts of the galaxy (around the same time a very similar technique was
presented by \citealt{fru06}, although these 
authors used the continuum blue light as a tracer of SF). 
This statistic then gives a value between 0 and 1 for each pixel
on the image, where 0 indicates a region of zero emission or sky, while a value of 1
indicates the centre of the brightest region of the galaxy. One can then use this statistic
to study the overall association of a certain type of object to e.g. the underlying
\ha\ emission of their host galaxies, where a higher mean pixel value indicates a
higher degree of association of that object type to the emission. We calculate pixel values
(dubbed NCR values in AJ08) for all
SNe with respect the \ha, 24\um, near-UV and far-UV emission and compare these
to the overall distributions previously found in AJ08.\\
One can also investigate SN progenitor and SF properties by investigating the 
radial distribution of different SNe within host galaxies. In AJ09 we presented results on the radial
positions of different SN types where we calculated the amount of host
galaxy light found within an ellipse or circle just including the SN position, 
normalised to the total galaxy. We 
measure these same radial values, referred to as Fr values, 
for all SNe, with respect to the $R$-band,
\ha, 24\um, near-UV and far-UV emission
found within the system. However, one can easily see when looking at Fig. 1 that 
this type of analysis is complicated in a galaxy system of this nature due to the 
irregular morphologies involved and problems defining a galaxy centre (the central point 
listed in NED and used in AJ09 is shown by the circle in Fig. 1). We discuss
these issues below, while also deriving alternative radial measurements with respect to 
the centres of the peak of the stellar mass.\\
Finally we derive metallicities for the local environments of each SN, from
the ratio of strong emission lines found within the host (or nearby) HII regions and assume these
as progenitor metallicities, using long slit optical spectra. To estimate metallicities from our spectra 
we follow the procedures outlined in A10 using the \cite{pet04} (PP04
henceforth) `empirical'
line diagnostics. These methods have the advantage of being relatively
insensitive to extinction, due to the close proximity (in wavelength) of the
emission lines used to derive metallicities. Using the same long slit observations we 
also extract numerous spectra along each slit (shown in
Fig. 2) to search for any possible metallicity gradients 
across the system.\\
In Fig. 3 we show example spectra extracted close to the position of the SN 2010P, marking the
emission lines that we use in our metallicity derivations.

\begin{figure}
\includegraphics[width=8.5cm]{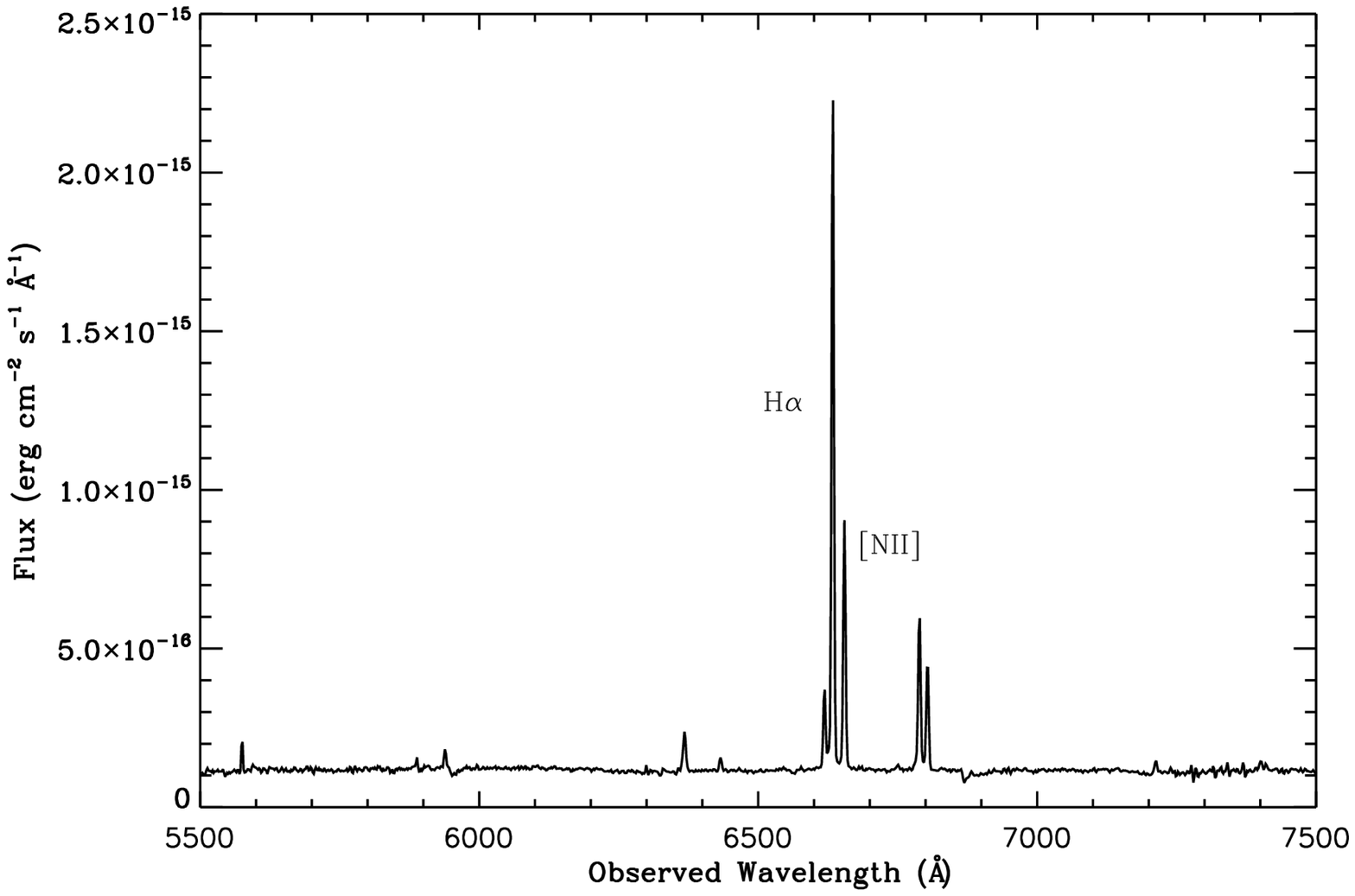}
\includegraphics[width=8.5cm]{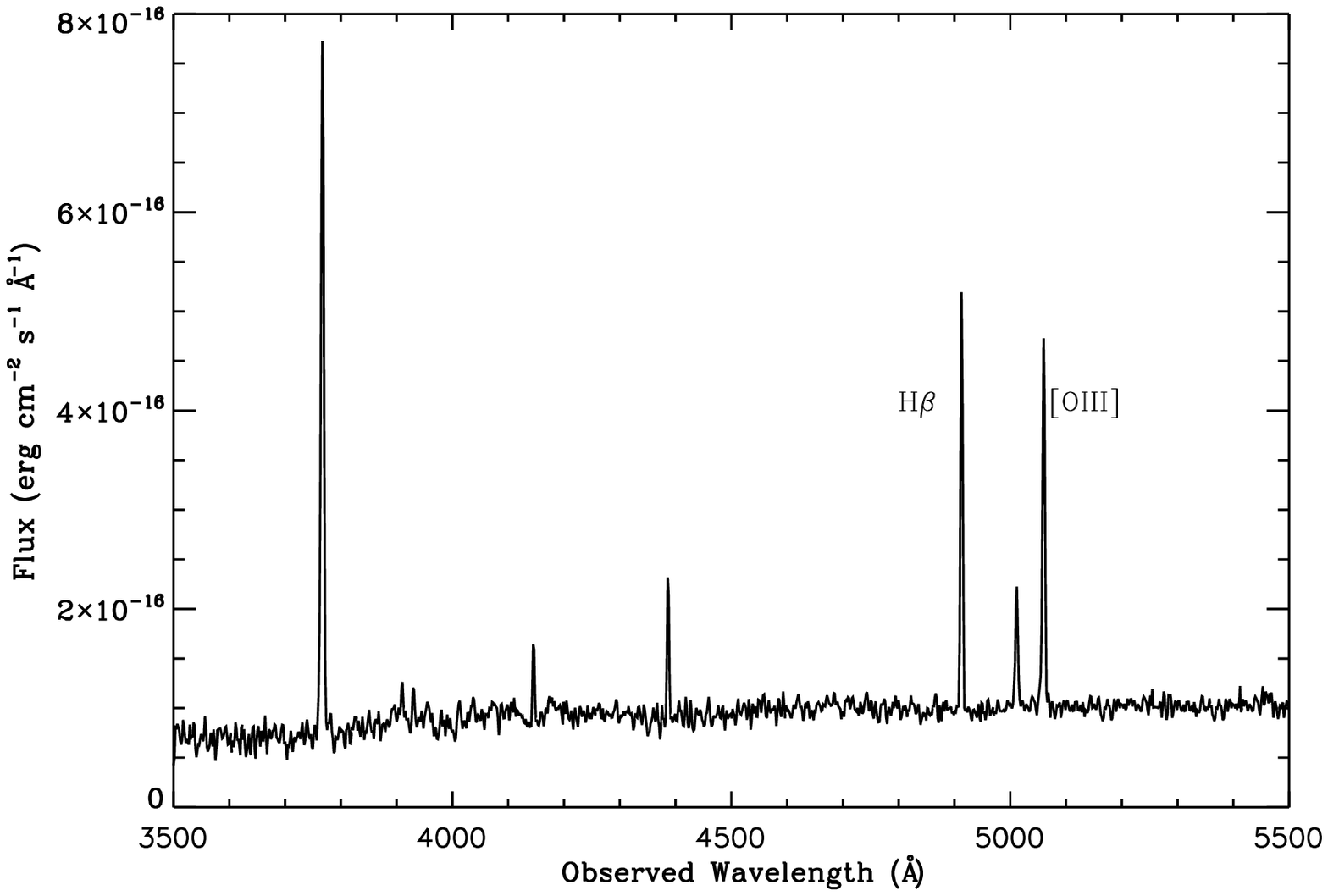}
\caption{Example spectra showing the red (top) and blue (bottom) arms of the ISIS spectra
extracted at the closest position along the slit to the position of the SN 2010P. The positions
of the emission lines used for the metallicity derivations are shown.}
\end{figure}

\section{Results}
\label{res}

\begin{table*}\label{results} \centering
\begin{tabular}[t]{ccccccccccc}
\hline
\hline
SN & Type & NCR$_{\textit{\ha}}$& NCR$_{\textit{Spitzer}}$ & NCR$_{\textit{nUV}}$& NCR$_{\textit{fUV}}$& Fr$_{\textit{R}}$&
Fr$_{\textit{\ha}}$& Fr$_{\textit{Spitzer}}$ & Fr$_{\textit{nUV}}$ & Fr$_{\textit{fUV}}$\\
\hline
1992bu & ? & 0.321 & 0.922   & 0.649 & 0.323 & 0.056 & 0.055 & 0.263 & 0.165 & 0.148 \\
1993G  & IIL & 0.064 & 0.193 & 0.194 & 0.098 & 0.464 & 0.744 & 0.761 & 0.534 & 0.514 \\
1998T  & Ib & 0.578 & 0.829  & 0.870 & 0.723 & 0.056 & 0.106 & 0.164 & 0.116 & 0.113 \\
1999D  & II & 0.054 & 0.122  & 0.263 & 0.407 & 0.560 & 0.849 & 0.990 & 0.899 & 0.895 \\
2005U  & IIb& 0.672 & 0.621  & 0.870 & 0.723 & 0.095 & 0.106 & 0.195 & 0.116 & 0.113 \\
2010O  & Ib & 0.421 & 0.559  & 0.508 & 0.285 &0.414 & 0.655 & 0.682 & 0.483 & 0.459 \\
2010P  & Ib/IIb & 0.406 & 0.662 & 0.563 & 0.414 & 0.095 & 0.106 & 0.195 & 0.138 & 0.129\\
\hline
\end{tabular}
\caption{NCR pixel values and fractional radial values for all SNe within Arp 299. In the first column
the SN name is given followed by the type classification. Then in columns 3 through 6 the NCR pixel values 
are listed for each SN with respect to the \ha, 24\um, near-UV and far-UV emission respectively. Then in 
columns 7 through 11 the fractional radial values are listed for each SN with respect to the $R$-band, \ha, 24\um, 
near-UV and far-UV emission respectively.}
\end{table*}

\subsection{Pixel statistics}
\label{pix}
The pixel statistics of each SN calculated from the different wavelength observations, together with the 
radial statistics are listed in Table 1.
The mean NCR \ha\ pixel value for the two non-IIb SNII is 0.059, i.e. both of these
SNe fall on regions with little \ha\ emission. For the SNIb (including the value for SN 2010P
at half weighting given the intermediate type classification of Ib/IIb), the mean is 0.481
and these SNe all fall on bright HII regions within the system. The mean value for the
SNIIb (again taking half weighting from SN 2010P) is 0.583. Finally the overall `stripped envelope' 
population has an \ha\ NCR mean value of 0.519.
Hence both the SNIb and SNIIb show a higher degree of association to the line emission
than the other SNII. This is consistent with the results from AJ08. 
We also note that the slightly higher degree of association of the 
SNIIb to the emission than the SNIb is consistent with that seen from the larger galaxy samples studied previously.\\
In a merging system such as Arp 299 there will be a high level of extinction due to dust. Therefore
\ha\ emission may not provide the most reliable indicator
of the distribution of SF. 
Emission in the mid-IR should be a sensitive tracer of such a dust-embedded SF component.
We therefore also calculate the NCR values for all SNe with respect to 
mid-IR emission at 24\um\ as imaged by \textit{Spitzer}. The mean NCR 24\um\ value for
the two non-IIb SNII is 0.158, 0.688 for the SNIb, and 0.635 for the SNIIb, with
a value of 0.668 for the overall `stripped envelope' population. Hence almost identical
trends are seen in the mid-IR as shown by \ha; the SNIb and SNIIb show a much higher
degree of association to the recent SF hence implying shorter pre-SN lifetimes and 
higher progenitor masses.\\
The pixel values derived for the near- and far-UV emission from GALEX images again show similar trends
to the above, with NCR values for the individual SNe being slightly higher than with respect
to the \ha\ emission. UV emission traces recent SF down to lower stellar masses and therefore
this is to be expected and shows that while the `normal' SNII do not fall on the brightest 
HII regions, they are still found to occur on regions
of recent SF, only those traced by lower mass stars than \ha\ line emission.\\
Almost exclusively in all tracers of SF the `stripped envelope' SNe
fall on higher pixel count regions of Arp 299 than the `normal' SNII (the only
anomaly is the NCR far-UV value for SN 1999D in comparison to that of
SN 2010P). Using this fact that all the 4 SNe Ib/IIb have higher NCR values
than the other SNe we calculate a $\sim$7\%\ probability of this occurring by
chance, if one assumes that in the true underlying distribution all SNe are
equally distributed on SF regions.

\begin{figure*}
\includegraphics[width=6.8cm]{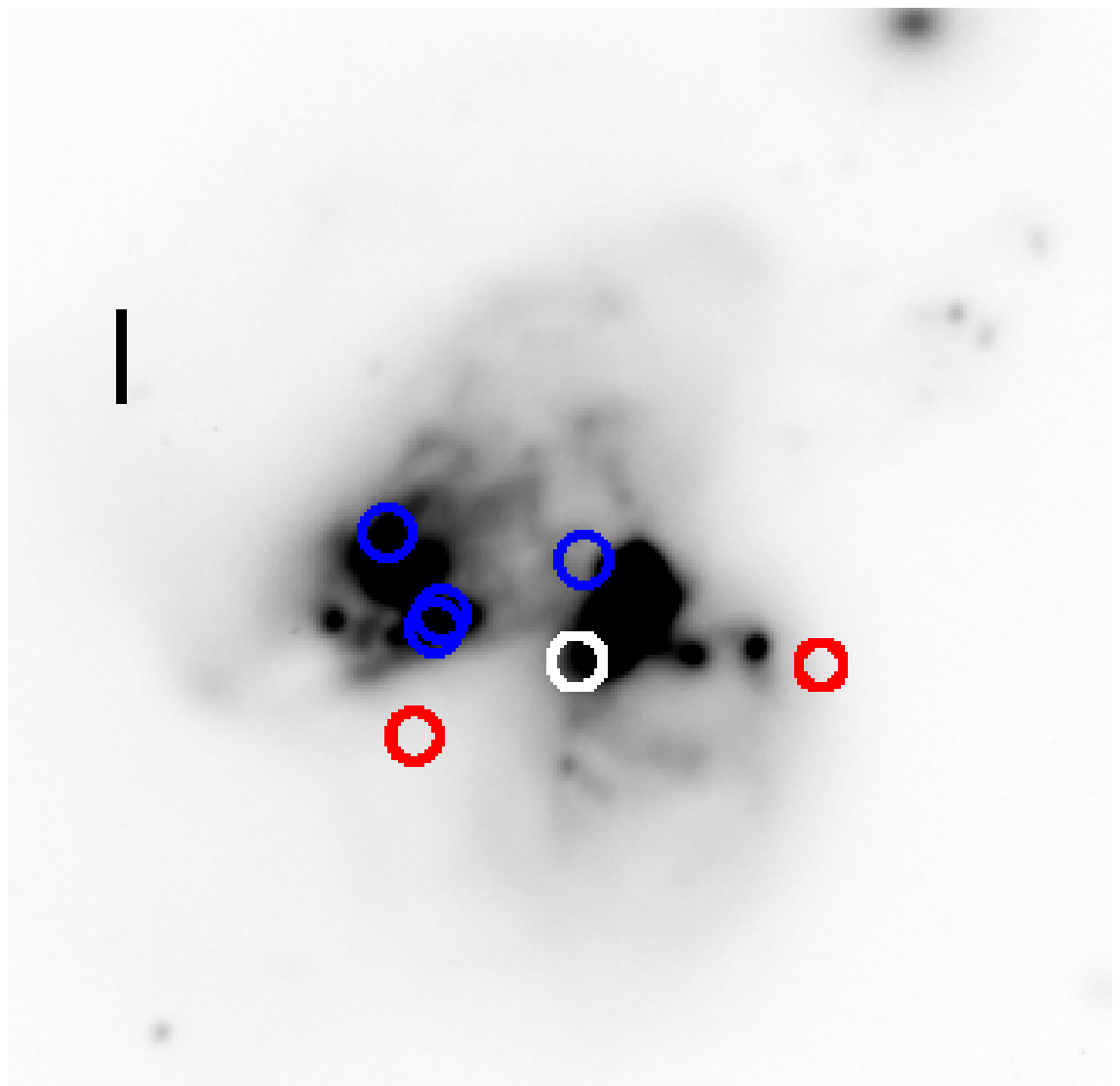}
\includegraphics[width=6.8cm]{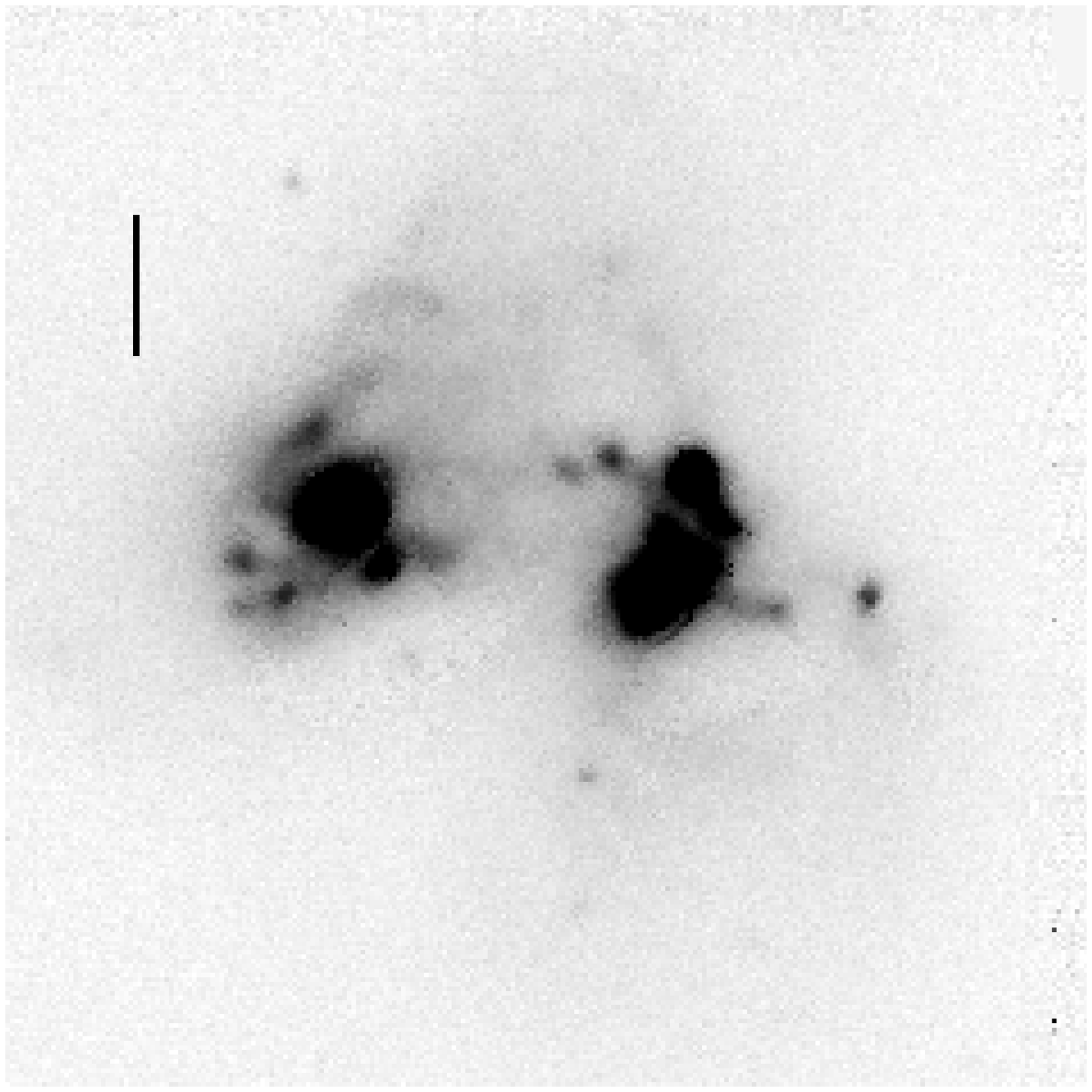}
\includegraphics[width=6.8cm]{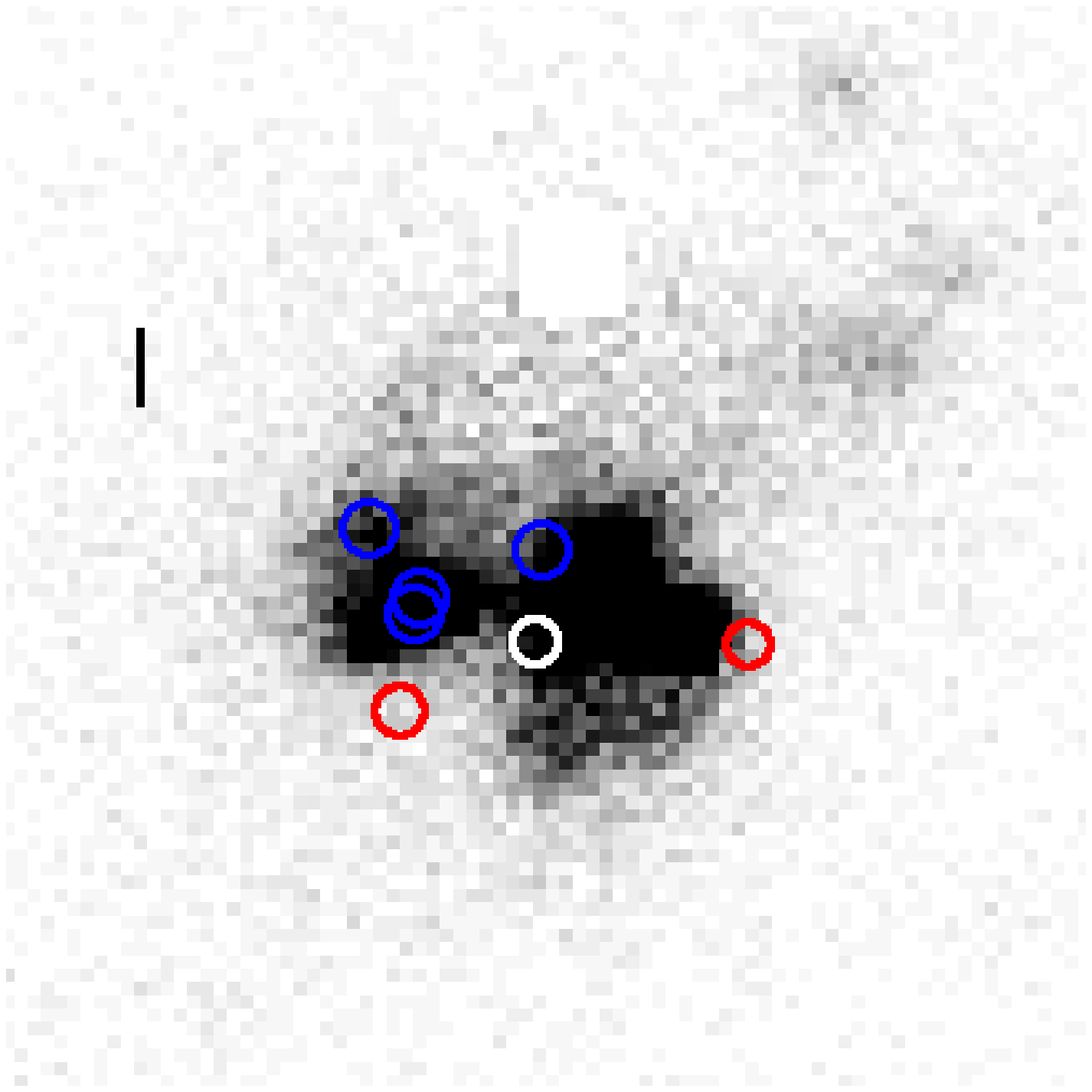}
\includegraphics[width=6.8cm]{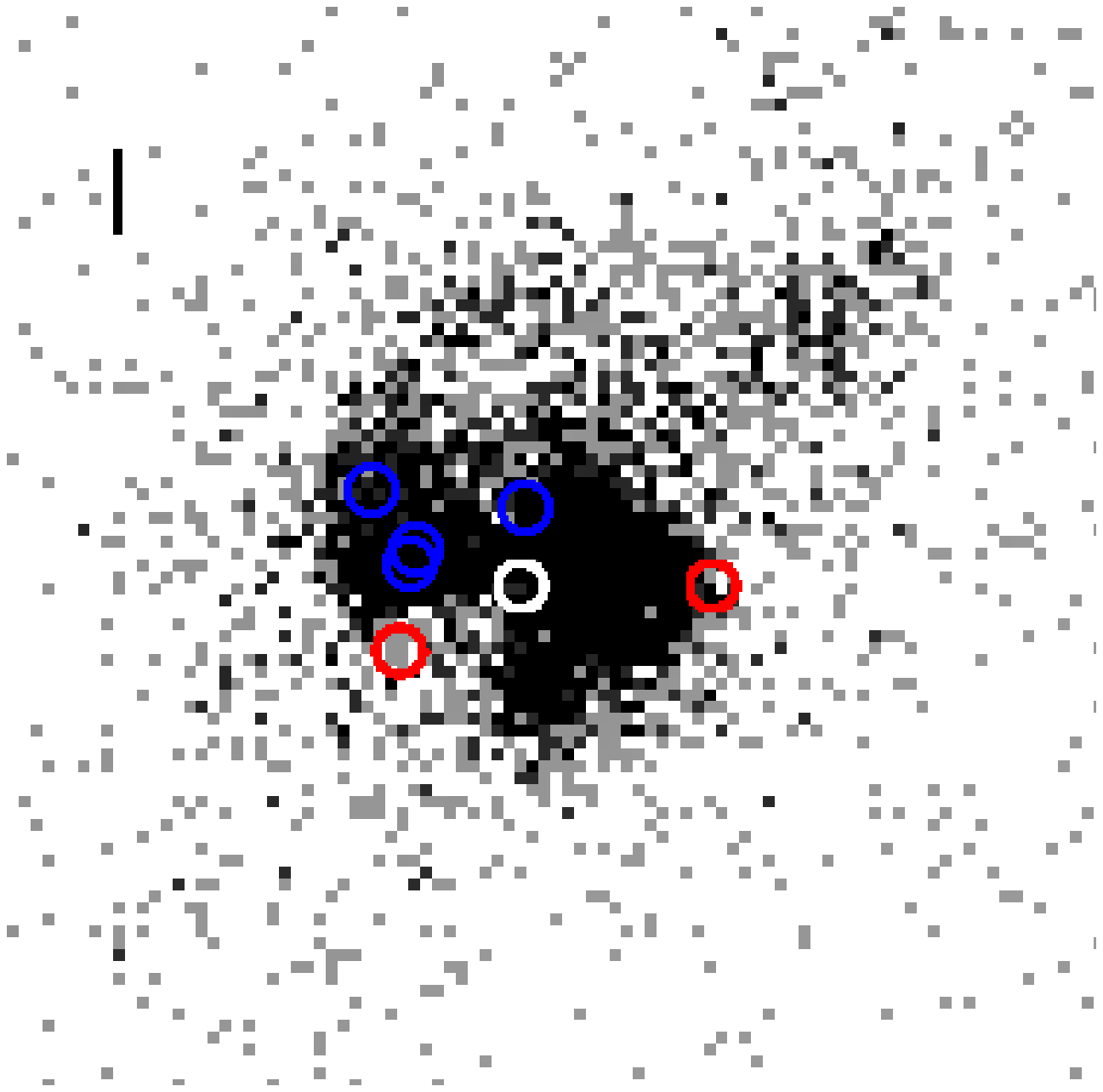}
\includegraphics[width=6.8cm]{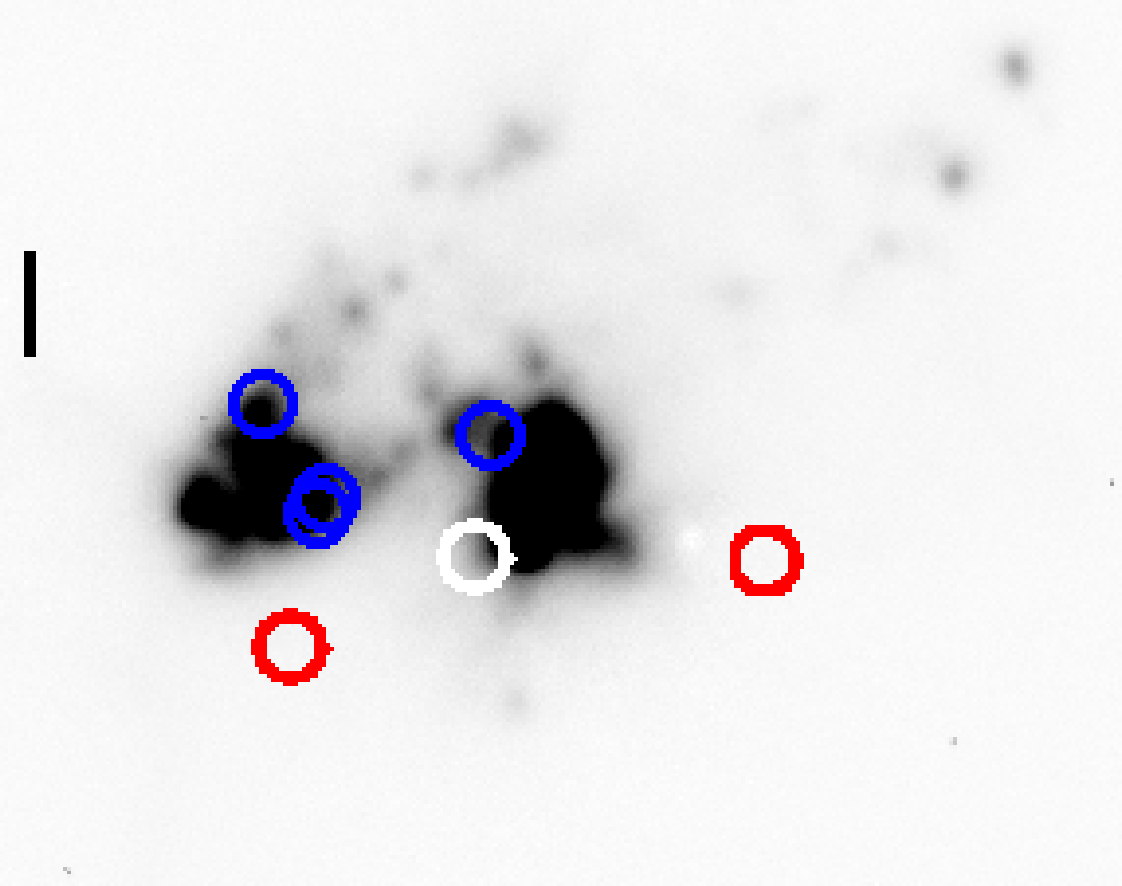}
\includegraphics[width=6.8cm]{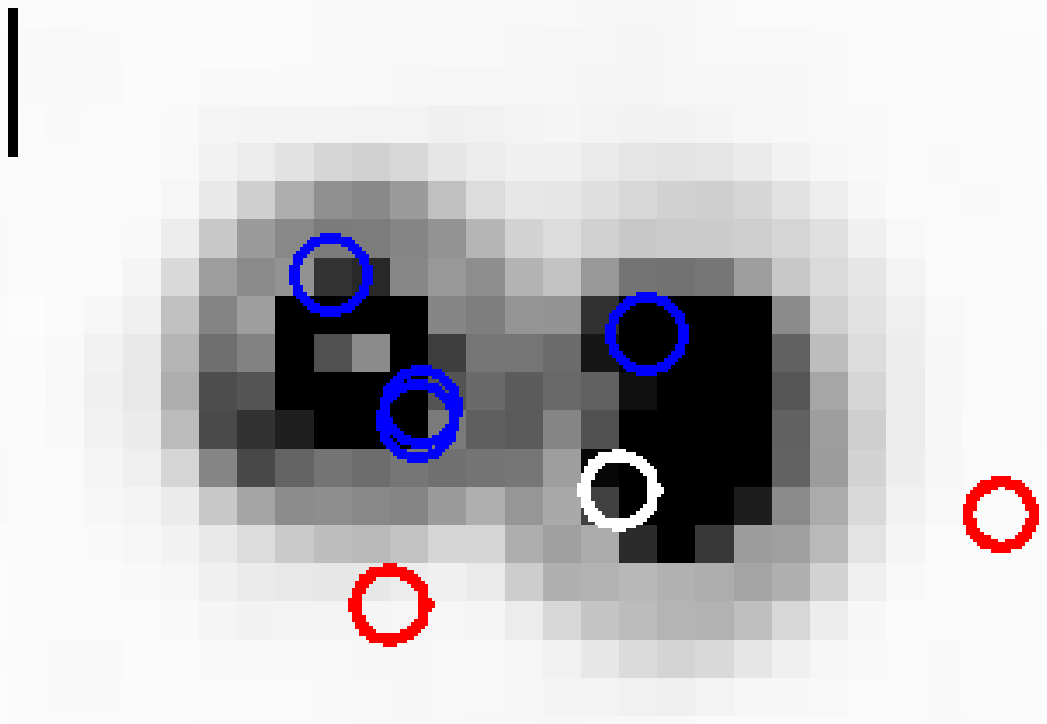}
\caption{Images of Arp 299 at different wavelengths. Top left shows
an $R$-band image with a $K$-band image shown top right (the $K$-band
image does not have the SNe positions marked as the image is not
astrometrically calibrated). In the middle panel the GALEX near-UV (left)
and far-UV (right) images are shown. In the bottom panel we show the \ha\ image
on the left and \textit{Spitzer} image on the right. On each image a scale 
bar shows a distance of 2 kpc, and the SNe positions are indicated by the coloured
circles: the `stripped envelope' SNe (Ib and IIb) are shown in blue
circles, while the `normal' SNII are shown in red. The unclassified SN
1992bu is shown in white.}
\end{figure*}

\subsection{Radial positions of SNe}
\label{rad}
We derive Fr values here with respect to Arp 299 central coordinates given in NED\footnote{Again we note the confusion on NED
with respect to naming conventions of this system discussed earlier. The
central coordinates used here are actually those obtained when searching NED for the coordinates
of NGC 3690. However, given where this puts the central point of the system (see Fig. 1), we
believe this to be the most sensible procedure. This is also consistent with the analysis in AJ09}. This
puts the centre midway between the two galaxy components (shown in Fig. 1), which may
be ambiguous, given that there are definitely (at least) two peaks in the emission of the merging 
system.
In AJ09 when dealing with interacting or irregular galaxies either
the two galaxies were treated separately or the central point of the $R$-band emission was used. 
Arp 299 is probably the only galaxy within that earlier sample where uncertainties in the central
location might significantly affect the conclusions drawn.
We present the Fr statistics for all SNe in the sample, and below proceed with additional analysis
to test the reliability of these results. The mean Fr$_{\textit{R}}$ for the 2 SNII is
0.512. For the SNIb the mean value is 0.207 while for the SNIIb the mean Fr$_{\textit{R}}$ value
is 0.095, with an overall mean `stripped envelope' value of 0.165. Hence both the SNIb and 
SNIIb are more centrally concentrated with respect to the $R$-band emission than the 
other SNII. In fact every individual SNIb and SNIIb is found closer to the central
point of the system
than the SNe 1993G and 1999D.\\
Again, very similar radial trends are found with respect to all other emission tracers 
used\footnote{The value in Table 1 for Fr$_{\textit{\ha}}$ of SN 1998T is slightly different 
from that listed in AJ09, this is due to a typo in that publication}.
This consistency of results across wavelength (here for the radial analysis and above for
the pixel statistics analysis) increases the robustness of the results presented.

\subsubsection{With respect to separate galaxy components}
\label{more_rad}
The merging nature of the system Arp 299 makes it almost impossible to 
separate the two components to allow for an unbiased study of the radial positions with respect to
the emission of the system. One way to investigate the radial positions further is to 
derive projected distances\footnote{We note that we do not attempt to
de-project these distances (or those used in the Fr analysis) 
due to the very uncertain nature of the inclinations etc of each
galaxy component. Therefore some of these quoted distances may be somewhat in
error. However, it would seem very unlikely that any de-projections would
heavily effect the overall results and conclusions} 
of each SN to the peaks of the stellar mass of each system,
as determined from the peak of the $K$-band emission from 2MASS \citep{skr06} images.
Here we determine distances for each SN from component $A$ (IC 694) and component \textit{B1} (brightest
peak of NGC 3690), following the naming conventions used in the literature (e.g. \citealt{geh83,alo00}).
These distances are listed in Table 2.\\
Associating each SN with the galaxy component in closest proximity we can again compare the values
for the different SN types. For the two `normal' SNII we find a mean distance of
3.82 kpc from their host nuclei. For the SNIb this value is 1.20 kpc while for the SNIIb the
mean is 1.33 kpc. The overall `stripped envelope' mean distance is 1.25 kpc. Hence while there 
are some small differences between the radial values here and those above (e.g. SN 2010O is 
considerably more central when this projected distance analysis is achieved), the overall
trends are again the same; the SNII are found further from the nuclei of the components of Arp 299
than both the SNIIb and the SNIb.\\
We also re-do the Fr analysis discussed above, centering the
radial apertures used to measure galaxy emission fluxes on the central
positions of NGC 3690 (\textit{B1}) and IC 694 ($A$). It is difficult to
assess where one galaxies' emission stops and the other starts due to the 
interaction of this system. Therefore here we include all of the emission of
the system in the analysis. While this will not give true Fr values for each
galaxy component, one can still look at the \textit{relative} Fr values
between the different SNe, and this is what we wish to test; do we still
see the trends from above that all the `stripped envelope' events are more
centrally
concentrated than the other `normal' SNII? The results of this analysis 
are shown in Table 3.
Again we see almost identical trends: almost exclusivley the Fr values are
lower for all `stripped envelope' events than for the other two SNe, and this
is seen across all wavelengths analysed.\\
As for the pixel stats above, we calculate the probability of this trend
occurring by chance, and the value is the same: $\sim$7\%, that 4 events from 6
are closer to the centres of the system assuming
that the underlying distribution is uniform across the whole galaxy.

The centralisation of SNIb and SNIIb over the other SNII is more difficult to interpret
than the pixel statistics above. If these SNe were occurring in a normal spiral galaxy then one
would probably infer that this centralisation were due to a metallicity effect. However, in 
a merging/interacting system such as Arp 299 any metallicity gradient is likely to be smoothed out 
by the inflow of gas; indeed we show this to be the case using our metallicity
derivations below. Other explanations include these central regions being too young to have produced many lower mass,
longer stellar lifetime SNII to date, or that in these central regions the IMF is biased
towards higher mass stars and hence is producing more SNIb and SNIIb over normal SNII. These 
final two points provide the main discussion points from 
this work and we return to them in detail in latter sections.\\

\begin{table}\label{results2} \centering
\begin{tabular}[t]{cccc}
\hline
\hline
SN & Type & kpc ($A$) & kpc (\textit{B1})\\
\hline
1992bu & ?  & 4.11 & 1.13\\
1993G & IIL & 3.37 & 4.69\\
1998T & Ib  & 0.98 & 3.77\\
1999D & II  & 8.71 & 4.27\\
2005U & IIb & 1.10 & 3.85\\
2010O & Ib  & 1.13 & 5.46\\
2010P & Ib/IIb & 3.74 & 1.79\\
\hline  
\end{tabular}
\caption{Distances (in kpc) of each SN from the nuclei of both NGC 3690 and IC 694.}
\end{table}

\begin{table*}\label{results6} \centering
\begin{tabular}[t]{cccccccccccc}
\hline
\hline
SN & Type & Fr$_{\textit{R}}$($A$)&
Fr$_{\textit{\ha}}$($A$)& Fr$_{\textit{Spitzer}}$($A$)&
Fr$_{\textit{nUV}}$($A$)& Fr$_{\textit{fUV}}$($A$)& Fr$_{\textit{R}}$(\textit{B1})&
Fr$_{\textit{\ha}}$(\textit{B1})& Fr$_{\textit{Spitzer}}$(\textit{B1})&
Fr$_{\textit{nUV}}$(\textit{B1})& Fr$_{\textit{fUV}}$(\textit{B1}) \\
\hline
1992bu & ?  & 0.353    & 0.359 & 0.413 & 0.344 & 0.281 & 0.131 & 0.117 & 0.046
& 0.120 & 0.096 \\
1993G & IIL & 0.277    & 0.320 & 0.381 & 0.260 & 0.235 & 0.511 & 0.426 & 0.469
& 0.705 & 0.734 \\
1998T & Ib  & 0.068    & 0.117 & 0.050 & 0.046 & 0.037 & 0.384 & 0.338 & 0.401
& 0.574 & 0.597 \\
1999D & II  & 0.841    & 0.942 & 0.763 & 0.919 & 0.920 & 0.511 & 0.426 & 0.490
& 0.692 & 0.724 \\
2005U & IIb & 0.068    & 0.117 & 0.070 & 0.059 & 0.050 & 0.423 & 0.359 & 0.401
& 0.613 & 0.645 \\
2010O & Ib  & 0.068    & 0.117 & 0.121 & 0.059 & 0.050 & 0.596 & 0.623 & 0.604
& 0.773 & 0.793 \\
2010P & Ib/IIb & 0.311 & 0.338 & 0.352 & 0.308 & 0.264 & 0.252 & 0.267 & 0.285
& 0.299 & 0.289 \\
\hline  
\end{tabular}
\caption{Fractional radial values (Fr) for all SNe with respect to the centres
of both IC 694 ($A$) and NGC 3690 (\textit{B1}). Columns 3 through 7 list the Fr
values with respect to $A$ and analysed with respect to the $R$-band, \ha, 
\textit{Spitzer}, near and far-UV emission respectively. While columns 8
through 12
list the same values but with respect to \textit{B1}.}
\end{table*}

\subsection{Environment metallicities of SNe}
\label{envZ}
In Table 4 we present the environment metallicities for all SNe discovered in
Arp 299\footnote{The extraction position for the environment of SN 1999D adopted here is
slightly different to that used in A10 following data reanalysis. This
results in small changes to the tabulated values, which do not affect the
overall conclusions}, together with their associated
errors\footnote{We note that our treatment of errors is different from
that in similar studies, e.g. \cite{mod11}, who treat the calibration
error as a systematic error on the PP04 scale and not for each individual
measurement. There is some evidence (e.g. \citealt{bre09}) that the individual 
measurements on e.g. the \textit{N2} scale are smaller then the quoted 0.18
dex. However, we note that adopting either treatment of the errors, the results
we obtain and the conclusions we draw below remain unchanged}. These metallicities are logarithmic oxygen abundances, relative
to hydrogen ([O/H]), which are derived using the PP04 \textit{O3N2} and
\textit{N2} emission line diagnostics\footnote{The overall errors on derived
metallicities quoted in Table 4 are slightly higher for the \textit{O3N2}
values and slightly lower for the \textit{N2} values than those published in
A10 (for SNe 1993G, 1998T and 1999D). This is due the wrong diagnostic
calibration errors being used for the SN metallicity errors in that publication. However, this does not
affect those published results or conclusions}. 
These are derived from spectra extracted at the closest position along the slit
to the catalogued SN coordinates. This means that in some cases the environment metallicity
is derived at a significant distance away from the actual SN position. These distances
are listed in column 6 of Table 4\footnote{Due to a typo in A10 the distance from the SN position
for SN 1993G was wrongly labeled as `SN'; here we list the correct
distance}.\\
One may therefore question whether these derived metallicities are
truly representative of those of the actual SN positions, or
even the `true' environment where the progenitor was formed (even
given the short stellar lifetimes of progenitor stars, it is still possible,
or even likely that stars will explode at considerable distances
from where they were formed; see \citealt{jam06} and A10 for additional
discussion). The environment metallicities of SNe 1998T and 2005U were both
derived very close to the catalogued positions, and hence we are confident
that these should be representative of the abundances of the actual
progenitors. 
For all other environment abundances were derived at $\sim$1 kpc or
more from the SN coordinates. Hence these values should be taken as a `best
guess' of the true progenitor metallicities. However, as many SNe occur away
from bright HII regions it is almost impossible to derive `true' emission line
metallicities (this is the case for SNe 1993G and 1999D in the present
study). Therefore we note this caveat to the values we present in Table 4, but
continue with the analysis, assuming these metallicities represent the
abundances near to the regions where these SNe exploded.\\
Looking at the environment metallicities presented in Table 4 we see that the 
inferred progenitor metallicities for all SNe discovered in Arp 299 are very similar,
with the largest difference between any two being 0.08 dex. For the 
two SNII we find a mean oxygen abundance, using the \textit{03N2} diagnostic of the 
PP04 scale of 8.55. For the SNIb this metallicity is 8.58 while for the SNIIb the mean is
8.54. For the overall `stripped envelope' population we derive a mean of 8.56. Hence all 
the environment metallicities for the different types are identical within the
errors.\\
We also compare our inferred progenitor metallicities with those
derived for larger samples in the literature. In A10 CC SN
progenitor
metallicities for 46 SNII (dominated by SNIIP), 10 SNIb and 14 SNIc were published. The mean
value for the SNII was 8.58, while for the SNIb the mean was 8.62. Hence
within the errors the values derived for the smaller samples here are
consistent with the mean values from A10; i.e. the inferred progenitor
metallicities for SNe in Arp 299 are neither metal rich nor metal poor when
compared to the overall populations studied previously. However, SNIb, SNIIb and
SNIc environment metallicities have also recently been published by
\cite{mod11} and \cite{lel11}. For the SNIb samples (\citealt{mod11} group the SNIIb in
this sample as well), these authors find mean metallicities (on the PP04
scale) of 8.41 \citep{mod11} and 8.52 \citep{lel11}. Hence with respect to
these studies, the SNIb and SNIIb metallicities we derive here
may
be considered slightly metal rich. However, the reasons for these
different metallicity estimates in different studies is not yet clear, and 
we note that while the above comparisons are interesting they do not affect
the overall conclusions we present in this paper. To resolve these issues
it is likely that we will have to wait for larger samples of
metallicities
to be published where all SNe are taken from `blind' SN searches.\\
Finally in
this section we compare the ratio of SN types found here with those found in 
the same metallicity range in A10. The mean of the 6 classified SNe in Arp 299
is 8.56 with a standard deviation of 0.05 dex. Taking all SNe that fall in
this metallicity range (8.50 to 8.61) from A10 we find a ratio of `normal' SNII to `stripped
envelope' SNe of 2.6. The A10 sample was a very heterogeneous group of SNe,
but this ratio agrees well with that seen in nearby galaxies, at all
metallicity
ranges (\citealt{li11}). However, the ratio of events we find in Arp 299 is 0.5. Hence, this
is further confirmation that differences in metallicity does not appear
a viable explanation for the types and distribution of SNe found in this
system.\\

The central parts of Arp 299 suffer from high
levels of extinction (e.g. \citealt{alo00}). Therefore one may worry that the metallicities derived for
circumnuclear regions (where
the majority of the SNe are found) may be significantly in
error. As discussed in A10, the extracted spectra of Arp 299 were not corrected for 
host galaxy extinction as the spectra were not taken at the
parallactic angle. However, as we use the PP04 diagnostics using emission lines
close in wavelength, this effect should only be
marginal. We check this
assumpition using the spectrum of the environment of SN 2010P (shown in
Fig. 3). \cite{mat10} and \cite{ryd10} reported an \textit{A$_v$} of $\sim$5 for SN
2010P, estimated from SN near-IR photometry and a deep optical spectrum.
Therefore we de-redden our SN enviroment spectrum by this amount (note that
the spectrum was not taken at the same position as SN 2010P, and therefore 
this comparison will be in error somewhat, however this is solely an example
of the effects of extinction). We then
re-measure line fluxes and re-estimate our environment metallicity values 
for SN 2010P. For the PP04 \textit{N2} diagnostic we measure a new abundance
of 8.69$^{+0.19}_{-0.19}$, 0.01 dex lower than the value reported in Table
4. While for the \textit{O3N2} diagnostic we measure a new abundance of
8.61$^{+0.14}_{-0.14}$, 0.02 dex higher than that reported in Table 4. Hence,
we conclude that even if there are large levels of environment extinction in
Arp 299, which we have not corrected for, the effect of these will be
negligible on our derived metallicities and hence the conclusions of this work.

\subsubsection{Comparison to literature values}
\label{complit}
Arp 299 has been extensively observed and studied in the literature. One such
study by \citeauthor{gar06} (2006, G06 henceforth) 
published emission line ratios for various regions across the system, using IFU observations.
Hence, as a sanity check we compare our derived environment metallicities with those calculated
from the emission line ratios published in G06. We determine the nearest emission region in G06
to our SN environment regions listed in Table 4. 
We can then compare our metallicities with those derived from the
flux ratios in G06. The results from this comparison are shown in Table 5 and
they illustrate that 
in general our derived environment metallicities agree extremely
well with those measured in nearby regions taken from G06, even though all of the regions compared are at least
0.2 kpc in separation. There are two cases where 
considerable metallicity differences are found. For SN 1992bu we find a metallicity difference of 0.14 dex. However, here
we associate the SN environment with the nucleus of NGC 3690, region \textit{B1}, where one may find much more extreme
conditions than outside this region, where our metallicity value is determined
(0.8 kpc away). Indeed, in 
regions \textit{B11} and \textit{B16} which are outside the main nucleus of NGC 3690 and only slightly further away
from our environment position than \textit{B1}, the metallicities one derives from the line ratios in
G06 are completely consistent with our values. Therefore we are not worried by this discrepancy. A  
difference (0.08 dex) is also found between the environment of SN 2010P and the value taken from the literature.
However, again if we compare our value with other nearby regions in G06 such as \textit{K6} or \textit{K7}
(which are separated to a similar degree to the separation between our environment and that taken from the literature),
we find a much better agreement.\\
In Table 5 we have listed the distances between the coordinates of the
extracted spectra we use for metallicity derivations, and the closest regions
in G06 used for the above comparison. As many of our metallicities are derived
from regions at significant distances from the catalogued SN positions one may
wonder whether some of the abundances derived from the emission line fluxes in
G06 may actually better represent those of the SNe currently studied. However,
calculating the projected distances between the G06 regions used above (and
indeed
other regions in G06 that may be closer to the SNe coordinates), we find that
all but one (SN 2010P, and here the difference is only 0.02 kpc) of our
extracted spectra are closer to the SN position than any regions studied in
G06.\\
Finally we note that even if we were to take the above literature
metallicities in place of our own values, the implications for the current study
would not change; there are no significant differences in the environment (and therefore progenitor) metallicities
between the different types of SNe found to occur in this interacting system.

\begin{table*}\label{results3} \centering
\begin{tabular}[t]{cccccc}
\hline
\hline
SN & Type & \textit{N2} & \textit{O3N2} & Extraction coordinates & Distance from SN position (kpc)\\
\hline
1992bu & ? & 8.76$^{+0.19}_{-0.19}$ & 8.60$^{+0.14}_{-0.14}$ & 11 28 31.33 +58 33 42.9 & 1.04 \\
1993G  & IIL & 8.56$^{+0.19}_{-0.19}$& 8.51$^{+0.14}_{-0.14}$ & 11 28 33.87 +58 33 37.4 & 1.61 \\
1998T  & Ib & 8.60$^{+0.18}_{-0.19}$ & 8.53$^{+0.14}_{-0.14}$ & 11 28 33.16 +58 33 43.7 & SN \\
1999D  & II & 8.75$^{+0.19}_{-0.19}$ & 8.59$^{+0.14}_{-0.14}$ & 11 28 28.77 +58 33 40.5 & 0.73 \\
2005U  & IIb & 8.60$^{+0.18}_{-0.19}$ & 8.52$^{+0.14}_{-0.14}$  & 11 28 33.26 +58 33 42.8& 0.10 \\
2010O  & Ib & 8.72$^{+0.19}_{-0.19}$ & 8.63$^{+0.14}_{-0.14}$ & 11 28 33.95 +58 33 45.2 &  1.36 \\
2010P  & Ib/IIb & 8.70$^{+0.19}_{-0.19}$ & 8.59$^{+0.14}_{-0.14}$& 11 28 31.81 +58 33 53.1 & 1.07 \\
\hline
\end{tabular}
\caption{Metallicity derivations for the closest HII region to the positions of each SN. 
In the first column the SN name is listed, followed by the type classification. In columns
3 and 4 the [O/H] abundances are given for each SN, derived from the PP04
\textit{N2} and \textit{O3N2} diagnostics, together with their associated
errors; a combination of the flux calibration error, the statistical error 
in the continuum at the position of the line measurements, and the calibration error
taken from PP04. In all cases this last calibration error dominates the overall metallicity
error. Then the coordinates are given for the extraction position (epoch J2000)
of the spectrum used to derive the environment metallicities. Finally the distance from the SN
explosion position to that of the extracted spectrum is listed (`SN' indicates that the extraction 
was achieved at the SN position).}
\end{table*}

\begin{table}\label{results5} \centering
\begin{tabular}[t]{cccc}
\hline
\hline
SN & G06 region & Metallicity difference (dex) & Distance (kpc)\\
\hline
1992bu & \textit{B1} & -0.14 & 0.8\\
1993G & \textit{G} & 0.00 & 1.0\\
1998T & \textit{A6} & 0.03 & 0.3\\
1999D & \textit{D3} & 0.01 & 2.1\\
2005U & \textit{A6} & 0.03 & 0.4\\
2010O & \textit{A1} & 0.04 & 0.2\\
2010P & \textit{K5} & -0.08 & 0.5\\ 
\hline
\end{tabular}
\caption{Metallicity differences between the SN environments (column 1) and 
regions taken from G06 (column 2). The metallicity differences are listed in
column 3 where a positive value indicates that the literature
value is higher than that derived in the current work, 
followed by the distance between the two regions in column 4. Note
that these regions are not necessarily the closest to the SN position, but are
the closest to the region used to infer the SN environment metallicities.}
\end{table}

\begin{table}\label{results4} \centering
\begin{tabular}[t]{cccc}
\hline
\hline
Extraction ID & \textit{N2} & \textit{O3N2} & Coordinates\\
\hline
1D & 8.71$^{+0.19}_{-0.19}$ & 8.62$^{+0.14}_{-0.14}$ & 11 28 33.73 +58 33 45.0 \\
2D & 8.66$^{+0.19}_{-0.19}$ & 8.59$^{+0.14}_{-0.14}$ & 11 28 32.98 +58 33 44.3 \\
3D & 8.69$^{+0.19}_{-0.19}$ & 8.61$^{+0.14}_{-0.14}$ & 11 28 31.95 +58 33 43.4 \\
4D & 8.79$^{+0.19}_{-0.19}$ & 8.58$^{+0.14}_{-0.14}$ & 11 28 30.44 +58 33 42.1 \\
5D & 8.75$^{+0.19}_{-0.19}$ & 8.56$^{+0.14}_{-0.14}$ & 11 28 29.83 +58 33 41.5 \\
6D & 8.75$^{+0.19}_{-0.19}$ & 8.60$^{+0.14}_{-0.14}$ & 11 28 29.02 +58 33 40.8 \\
1T & 8.69$^{+0.19}_{-0.19}$ & 8.55$^{+0.14}_{-0.14}$ & 11 28 33.98 +58 33 36.5 \\
2T & 8.65$^{+0.19}_{-0.19}$ & 8.54$^{+0.14}_{-0.14}$ & 11 28 33.61 +58 33 39.7 \\
3T & 8.59$^{+0.18}_{-0.19}$ & 8.55$^{+0.14}_{-0.14}$ & 11 28 33.20 +58 33 43.4 \\
4T & 8.70$^{+0.19}_{-0.19}$ & 8.61$^{+0.14}_{-0.14}$ & 11 28 32.61 +58 33 48.6 \\
5T & 8.69$^{+0.19}_{-0.19}$ & 8.59$^{+0.14}_{-0.14}$ & 11 28 32.04 +58 33 53.6 \\
6T & 8.78$^{+0.19}_{-0.19}$ & 8.56$^{+0.14}_{-0.14}$ & 11 28 31.30 +58 34 00.2 \\
7T & 8.69$^{+0.19}_{-0.19}$ & 8.59$^{+0.14}_{-0.14}$ & 11 28 28.99 +58 34 20.6 \\
\hline
\end{tabular}
\caption{Metallicity values for the various slit positions indicated in Fig. 2. The number before
the letter `D' or `T' indicates the extraction position along the slit with the number
one extraction being that furthest to the left (east) of the galaxy image in Fig. 2. For 
each slit position the environment abundances are listed together with
their associated errors, followed by the coordinates of the extraction position (epoch J2000).}
\end{table}

\subsection{Metallicity gradients in Arp 299}
\label{grad}
Using the slit positions as shown in Fig. 2 we can also
investigate whether there are any metallicity gradients across the
system. Spectra were extracted at the positions shown in Fig. 2 and metallicities 
were derived in the same way as above for the environment HII regions. These metallicities
are listed in Table 6 where the numbers followed by `D' refer to the SN 1999D slit position
(centred close to the position of SN 1999D with a position angle of 83.4 degrees east of north)
and those referred to as numbers followed by `T' refer to the SN 1998T slit position (centred
close to the position of SN 1998T with a position angle of 131.5 east of north).\\
One can see in Table 6 that we find no gradients across the system, and indeed very little variation in metallicity
from environment to environment. This is completely consistent with the above SN environment metallicities
and reinforces the finding that nearly all environments within the interacting system Arp 299 have
very similar gas-phase metallicities. This is also consistent with recent work \citep{kew10} that has found that
metallicity gradients within interacting galaxy pairs are much shallower or even non-existent 
when compared to normal isolated spirals.

\subsection{Relative rates of SNe in Arp 299}
\label{rates}
One of the motivations for this case study was the suggestion that the SNe types 
that have been discovered in the system are dominated by the `stripped envelope'
CC SNe (i.e. the Ib, Ic or IIb). In this section we analyse this hypothesis. 
6 SNe with classifications have been found in the system; one SNII (no
further classification), one SNIIL, one SNIIb, two SNIb and an event classed as only Ib/IIb.
Above we have considered three groups; the `normal' type II (in which we
include the SNIIL), the IIb and the
Ib, while also considering the `stripped envelope' SNe as an overall group.
Calculating a ratio of `normal' SNII to `stripped envelope' events we find a value
of 0.5 in Arp 299. We can then compare this value to nearby SN rate
measurements. The most recent rates were measured and published by the LOSS
group \citep{li11}. From their measured rates we find a ratio of $\sim$2; a factor of 4
higher than that found in Arp 299.\\
To calculate the probability of finding the observed SN type distribution in
Arp 299 we use a Monte Carlo analysis to draw 6 SNe randomly from the overall
nearby CC SN rates published by \citeauthor{li11} (as percentages of the CC SN population we derive: 52\%\ SNIIP, 7\%\ IIL,
9\%\ IIb, 7\%\ IIn, 7\%\ Ib and 18\%\ Ic). We can then derive the
probability of detecting a ratio of `normal' type II SNe to `stripped
envelope' events 
equal to or less than the value of 0.5 we find in the
current analysis. Note, here we include the SNIc in the `stripped envelope'
group for the probability analysis, even though there are no SNe of this type
detected in the galaxy; this is to be consistent with the separation of SNe
throughout the above analysis into those the in `stripped envelope' class and
those that are not (we also note there is some
ambiguity in the clasification of some SNe between the Ib and 
Ic subtypes; \citealt{lel11}). We derive a probability of $\sim$10\%\ that one would find a
ratio equal to or less than that found here. Hence there is marginal
statistical evidence that the number of `stripped envelope' events is higher
in this merging system than that found in nearby galaxies. (If we were
not to include the SNIc in the probability analysis, and derive the likelihood
of finding a ratio of Ib and IIb events to all others of less than or equal
to 0.5, this probability drops to 0.8\%). However, we note that SNIIL 
(SN 1993G) are relatively rare events and one may consider these `stripped
envelope' events when compared to SNIIP (i.e. SNIIL 
have smaller hydrogen envelopes at the time of explosion). Hence if we
were to include this in the probability calculation, we find only a
5\%\ chance that if we were to draw a sample of 6 SNe randomly from those
found
in nearby galaxies we would find a ratio of 1:5 of `normal' SNII to other
CC types. Finally we note that even this probability may be an overestimation,
given the undistinct sub-type classification of SN 1999D.\\
Within Arp 299 there is one unclassified SN; 1992bu. To estimate the
error on the above chance probability, we can re-run our Monte Carlo simluation assuming that 
SN 1992bu is either a `stripped envelope' or `normal' event. We
find a probability of $\sim$5\%\ if we include the SN in the former, and 
a probability of $\sim$19\%\ if we were to include the SN in the latter (that
we observe the detected SN type distribution by chance assuming underlying
SN rates found in the local Universe). In
section 6.4 we argue that given the environment of SN 1992bu (high degree
of association to SF tracers, found to occur centrally), 
it is more probable that the SN was of the `stripped envelope' 
type. Hence if we take this to be true, the claim of a relatively higher rate
of `stripped envelope' events to `normal' SNII in Arp 299 would be strengthened.


\section{Discussion}
\label{diss}
In the previous sections we have presented various analyses investigating the relative
numbers of different SN types, together with the positions at which they are found
within their host galaxy system. Overall we find that
the ratio of `stripped envelope' CC SNe to `normal' type II events is higher
than that found in samples of nearby galaxies, and that the former are much more
centrally concentrated within the system than the latter. In the following sections we
discuss the possible implications of these findings, both on SN progenitor properties and the 
nature of the SF found within this system.

\subsection{The environments of SNe within Arp 299}
\label{environ_diss}
The `stripped envelope' SNe are all found to occur more centrally with respect to the stellar
populations traced by different wavebands, and also with respect to the peaks in the stellar
mass distribution of the two galaxy components than the other SNII. Usually such centralisation would be
attributed to a metallicity affect. However, as we show above, no metallicity gradients exist
within this system and abundance derivations close to the sites of each SN position show
no preference for the `stripped envelope' events to occur in higher metallicity 
environments. These SNe are also found to occur within bright HII regions within the system, while
the other SNII are found further from such high mass star tracers. This implies that the latter
have arisen from lower mass progenitors, consistent with previous results looking at large 
samples of CC SNe in nearby galaxies (AJ08). 

\subsection{Starburst age effect}
\label{starage}
One possibility that may explain the relatively high occurrence of `stripped envelope'
events is that the majority of the SF within the system is too young for the
`true' SN rates (assuming a non-varying IMF) to be observed. This would be consistent with the
very young ages of the most recent SF episodes within system found by G06.
These authors generally find ages for the HII regions found within Arp 299 between 3-7 Myr. 
If SNIbc arise from single star progenitors then they are likely to arise from
stars more 
massive than 25\msun\ (e.g. \citealt{heg03,mey05}). From the models of \cite{mey05} this corresponds
to a pre-SN progenitor age (summing the hydrogen and helium burning timescales) of $\sim$8 Myr. Hence this
is reasonably consistent with the interpretation that the dominant current SF within the merging event is
on timescales where only stars of masses consistent with those for SNIbc and indeed the overall
`stripped envelope' class have had sufficient time to explode as SNe (it is indeed possible that
we are observing the system at such a specific time when all of the most massive stars have already exploded 
as SNIc and hence we do not observe any of these SNe, but SF within the system is not old enough to be 
dominated by SNIIP explosions and hence we find an overabundance of SNIb and
SNIIb).\\ 
If this is indeed the case then 
the relative numbers of SNe within Arp 299 provides \textit{independent} 
(from that presented in AJ08) observational evidence that SNe of type Ib and IIb arise from shorter lived
and hence more massive progenitors than other `normal' type II
events. 
This
line of reasoning also gives further evidence for the progenitor mass
sequence suggested in AJ08. Here, the reason we do not observe any
SNIc (as outlined above) is because the most massive stars have already
exploded as SNIc. The lack of SNIc and the overabundance of
SNIb/IIb then suggests a progenitor mass sequence
from SNII arising from the lowest mass stars, through
the SNIb and SNIIb, and finally the SNIc arising from highest mass progenitors.
(We make an important point here: this does
not necessarily favour single star progenitors for these events over binary progenitor scenarios. One can
easily envisage a scenario where some, possibly the majority of these events arise from binaries, but the 
components of those binaries are indeed more massive and have shorter lifetimes than those that explode as SNIIP or SNIIL events.)
This is also consistent with the work of \cite{nel10}, who tentatively concluded that the progenitor of
the SN 2010O (type Ib) was a massive Wolf-Rayet star from analysis of \textit{Chandra}
pre-explosion images.\\
There are however a number of caveats to this interpretation. Firstly, 
the ages published for SF regions studied by G06 represent the ages of the most recent extreme SF. 
The above interpretation would require either that before these episodes there was very little SF taking place, or that
the SF rate over the last few Myr is \textit{much} higher than that previously meaning that SNe produced from these
recent events will dominate over those produced by older but longer duration SF episodes. However, the interaction/merging
of the two components of Arp 299 has probably been active for at least a few 100 Myr and therefore it is difficult to see why
the SF from the last few Myr should be so dominant. Indeed, we do find 2 SNe classified as SNII which we presume to have
lower mass progenitors, and therefore there is ample evidence for significant
SF over timescales longer than that found in previous
studies, and that required to produce the high number of `stripped envelope' SNe seen in this study.
It is also not clear how this interpretation would explain the centralisation of the `stripped envelope' SNe
over the other 2 SNII. One could envisage a scenario whereby the majority of the SF is happening towards the central 
parts of the system, then the longer lived progenitor (those of the SNII) have time to drift
away from the central regions before exploding. However, this would require that the SF within these central regions
has been active and continuous for timescales longer than those quoted above, and indeed this would then not explain
why there are so many of the `stripped envelope' SNe with respect to the others.\\
In conclusion to this section, it is very possible that the age of the recent extreme SF currently occurring 
in Arp 299 could explain some of the results we present in this work. However, for this to be true a number of 
(in our opinion) unlikely coincidences would have to be at play. Therefore we argue that there are probably other
factors affecting the observed distribution of SNe in Arp 299 which we now discuss.

\subsection{Comments on IMF interpretation}
\label{IMF}
In H10, we demonstrated that SNIbc show a striking degree of central
concentration in disturbed host galaxies, and that the central regions
of these systems produce a remarkably small fraction of SNII.
Disturbed galaxies are known to experience nuclear starbursts, with
Arp 299 being a classic example, and several studies (see H10 for
further details) have inferred that these starbursts are likely to
exhibit a modified stellar initial mass function (IMF), biased to
produce relatively more high mass stars than a `standard' IMF
(e.g. Salpeter 1955). Such a modified IMF gives a natural explanation for an
enhanced fraction of `stripped envelope' SNe, if the latter result from higher-mass
progenitors than SNII, as was found by AJ08.  Indeed, Klessen et
al. (2007) suggest a specific mechanism for modifying the IMF in
starbursts which suppresses or completely prevents the formation of
all stars below a given mass limit, which could explain the deficiency
of SNII found in disturbed galaxy centres. The predictions of this
model for CC SN fractions generally will be explored in a later paper
(James et al. in preparation), but it is interesting to note that the
Klessen model requires a high-temperature intergalactic medium, with a
consequent raising of the Jeans mass, a prediction that seems to be
strongly confirmed in the case of Arp 299.  The IRAS
60$\mu$m/100$\mu$m flux ratio for Arp 299/NGC 3690 is 1.01,
approximately twice the typical value for galaxies in the Revised IRAS
Bright Galaxy Sample (BGS) of Sanders et al. (2007); only $\sim$4 per cent
of galaxies in the BGS have a flux ratio exceeding unity.  Thus
Arp 299 is exactly the type of galaxy where the effect predicted by
Klessen et al. (2007) is likely to be seen, giving a plausible, though
clearly non-unique, explanation of the unusual SN type and spatial
distributions found in the present work.

\subsection{The environment and progenitor of SN 1992bu}
\label{unclass}
SN 1992bu has no type classification in the literature. However, given the 
above results and discussion we can make some tentative claims as to its possible progenitor and
hence SN type. 
SN 1992bu falls on a bright SF region (particularly evident in the 
high \textit{Spitzer} NCR value; see Table 1) and is also very close to both the centre of NGC 3690 and the 
overall centre of the system (Tables 1, 2 and 3). Hence, given this information
we conclude that 
the environment of SN 1992bu is consistent with the SN being of `stripped envelope' classification (i.e. 
arising from a higher mass progenitor). 
There are obviously huge uncertainties in this conclusion, and we 
stress that our overall discussion and conclusions are independent of the 
adopted classification of SN 1992bu.

\subsection{Caveats}
\label{caveat}
A major caveat that must be discussed in any study such as that presented here is how
extinction may affect any observations and measurements that are made. We noted earlier
that there are many regions within Arp 299, especially in the central component peaks, that
are very heavily extinguished by dust \citep{alo00}. Therefore many SNe will go undetected
in optical or even near-IR SN searches. However, this is only an issue for the current investigation and its implications 
if there is a strong bias in SNII being missed more often than other types because of extinction.
SNIIP are generally intrinsically fainter than other CC SN \citep{ric02,ric06} and therefore may 
be harder to detect especially in the highly extinguished regions found in Arp 299 (in the central regions
of the galaxy components within Arp 299 there is also likely to be problems detecting SNe due
to high surface brightness, although in the case of this merger it is probable that extinction due to 
dust is the dominant detection problem). However, 
these SNe magnitude differences are reasonably small and 
to find the observed distribution of SNe in Arp 299 this effect would have to be precisely 
tuned so that \textit{zero} SNIIP (by far the most dominant CC SN type in nearby galaxies) were
detected in the central regions (or indeed anywhere in Arp 299), while significant numbers of
other `stripped envelope' events are detected. Arp 299 is currently being monitored at a
number of different wavelengths (e.g. in the radio by \citealt{per09}) and therefore given the 
high current SF rate in Arp 299 one expects additional SNe to be observed over the coming years. 
Any additional events will go some way to validating (or invalidating) the interpretations put forward
in the current paper.

\section{Conclusions}
\label{con}
In this work we have presented a case study of the relative number of SNe
and their positions within Arp 299. 
We find an unusually high fraction of 'stripped envelope' SNe relative to 
other type II events ($\sim$10\%\ chance probability), and the 
former are all more centralised within the system than the latter 
($\sim$7\%\ chance probability). Taking these findings together, we can draw two 
distinct - but non-mutually exclusive - interpretations.
Firstly these results can be explained by the
system being dominated by a very young burst of SF. Here the system is too young for the
observed relative SN rate to match the `true' rate expected from a standard IMF.
This interpretation gives additional, independent observational evidence that both
SNIb and SNIIb arise from shorter lived and more massive stars than those coming
from other `normal' SNII. Secondly if (from previous research) we assume that SNIb and SNIIb arise 
from more massive stars, then this implies that within the central regions of the
Arp 299 system the IMF is biased towards the production of higher mass progenitor stars.
While overall we believe that it is likely that both of these interpretations are at play
at some level, our preference is that the IMF interpretation is the dominant one as
this most naturally explains the observations of Arp 299 and its SNe without resorting to too many
coincidences and special circumstances.

\section*{Acknowledgments}
We thank the referee Seppo Mattila for constructive comments. We also
thank Maryam Modjaz, Giorgos Leloudas and the MCSS group for useful discussion.
J.A. acknowledges fellowship funding from FONDECYT, project number 3110142 and
partial support from the Millennium Center for Supernova Science through grant P06-045-F 
funded by ``Programa Bicentenario de Ciencia y Tecnolog\'ia de CONICYT'' and ``Programa Iniciativa Cient\'ifica Milenio de MIDEPLAN''.
This research
has made use of the NASA/IPAC Extragalactic Database (NED) which is operated by the Jet Propulsion Laboratory, California
Institute of Technology, under contract with the National Aeronautics and
Space Administration and of data provided by the Central Bureau for Astronomical Telegrams. 
We also acknowledge the usage of the HyperLeda database
\bibliographystyle{mn2e}

\bibliography{Reference}

\label{lastpage}

\end{document}